\documentclass[conference]{IEEEtran}
\IEEEoverridecommandlockouts
\usepackage{cite}
\usepackage{amsmath,amssymb,amsfonts}
\usepackage{algorithmic}
\usepackage{graphicx}
\usepackage{textcomp}
\usepackage{xcolor}
\usepackage{url}
\def\BibTeX{{\rm B\kern-.05em{\sc i\kern-.025em b}\kern-.08em
    T\kern-.1667em\lower.7ex\hbox{E}\kern-.125emX}}

\begin{document}

\title{"Optimizing Bus Route Selection for University of Pittsburgh Students: A Comparative Analysis of Ridership, On-Time Performance, and Travel Distance"\\
}

\author{\IEEEauthorblockN{Logan Warren (Lead Author)}
\IEEEauthorblockA{\textit{Pittsburgh, US} \\
\textit{law234@pitt.edu}\\
}
\and
\IEEEauthorblockN{Jerek Stegman (Contributor)}
\IEEEauthorblockA{\textit{Pittsburgh, US} \\
\textit{jjs252@pitt.edu}\\
}
}

\maketitle

\begin{abstract}
Abstract- This paper describes a research project that looks into the topography of Pittsburgh’s bus routes using geographical data obtained from The Western Pennsylvania Regional Data Center. This research focuses on comparing bus route usage to find the route with the least occupants, and bus routes average on time performance, to get from the University of Pittsburgh to the most centralized bus stop in Downtown Pittsburgh. We will use computational software, such as R-Studio to find correlation between our data, and come to a conclusion on the least crowded route, that will have the fewest stops, and the fastest time. This research was done for INFSCI 310 - Computation in Information Science, a class at the University of Pittsburgh’s School of Computing and Information (SCI).
Keywords - Bus routes, topography, geographical data, R programming language
\end{abstract}

\begin{IEEEkeywords}
"public transportation, bus routes, University of Pittsburgh, data analysis, ridership, on-time performance, R programming language, Python, Pandas."
\end{IEEEkeywords}

\section{Introduction}
The motivation behind our research project was to determine the best route for traveling from the University of Pittsburgh to the central-most bus stop destination in Downtown Pittsburgh. Our conclusion is based on the analysis of four key datasets: Transit Stop Usage by Route, Monthly Average Ridership by Route \& Weekday, Pittsburgh Regional Transit Transit Stops, and Monthly On-Time Performance (OTP) by Route. These datasets were chosen because they provide a comprehensive measure of bus usage, timings, and stop locations, allowing us to identify the most efficient route between the two points. With this information, we are looking to find the best route to take from the University to Downtown, focusing on 3 main factors. First, we want to find which bus routes have the least amount of ridership. Second, we want to use our data to determine the shortest route to a central point in downtown. Lastly, with the On Time Performance data, we will consider the buses' punctuality to gauge our quickest route.

The public transportation system in Pittsburgh plays a crucial role in the city's daily life, with many residents, including students, relying on it for their commute. The Port Authority of Allegheny County operates a network of buses, light rail, and inclines that serve the Pittsburgh metropolitan area. The University of Pittsburgh is well-served by the public transit system, with numerous bus routes passing through the campus and connecting it to other parts of the city. However, navigating this extensive network can be challenging for newcomers, particularly students who are new to the city and its transportation system.

According to 2022 data, approximately 40\% of undergraduate students at the University of Pittsburgh are from out-of-state, which is roughly 7,000 students. Being new to the city and its bus system can be confusing and challenging for these students. As part of this group, we can attest to the difficulties we faced in finding our way around Pittsburgh and reaching our destinations on time. The University of Pittsburgh is located just outside of walking distance from the downtown area, so when students need a mode of transportation, they often turn to the public bus system. However, many students find themselves unsure about which bus to take, which route will get them to their destination the quickest, and which bus will be the least crowded.

Determining the best bus route from the University of Pittsburgh to Downtown Pittsburgh is essential for student travel and success. Our goal is to identify the optimal route that has the fewest stops, arrives on time, and caters to students' various needs, such as weekend outings, internships, and commuting to and from school. This information will provide students with a reliable and efficient mode of transportation option.

To analyze and compare the data from our chosen datasets, we will utilize Python and Pandas for data manipulation and analysis. Pandas is a widely-used data manipulation library for Python. We will also focus on using R, a programming language for statistical computing, for additional data manipulation and to create visual representations of our findings. By employing these two programming languages, we will generate various metrics that will help us determine the best route for traveling from the University of Pittsburgh to Downtown Pittsburgh.

\section{Project Goals}

The primary goal of this project is to analyze four datasets from the Western Pennsylvania Regional Data Center and identify the most efficient bus route for travel from the University of Pittsburgh to the central-most bus stop in Downtown Pittsburgh. This objective is essential as it provides students with a reliable and convenient transportation option that they can take to downtown for various reasons, taking into consideration the three critical factors previously mentioned.
Our research aims to answer four main questions:
\begin{itemize}
\item Which route reaches the central point of Downtown Pittsburgh the quickest, considering the frequent stops?
\item Which bus route has the lowest ridership, ensuring a quick and comfortable ride?
\item Which bus route has the highest on-time performance?
\item What is the best overall route for traveling from the University of Pittsburgh to Downtown Pittsburgh based on the information gathered from the previous questions?
\end{itemize}
Based on our research, we anticipate that the fastest route to central Downtown from the University of Pittsburgh will likely follow Boulevard of the Allies, as it offers a direct path from the university to downtown. As for the route with the lowest ridership, we expect a bus that does not travel down Fifth Avenue and instead traverses a different road or a bus-only lane, avoiding major stops around campus where riders frequently board.

In order to achieve our project goals, we will implement the following steps:
\begin{enumerate}
\item Obtain and preprocess the datasets from the Western Pennsylvania Regional Data Center.
\item Utilize Python and Pandas for data manipulation and initial analysis.
\item Use R for further data manipulation and creating visual representations of our findings.
\item Analyze the datasets and compare the bus routes based on the three critical factors: ridership, on-time performance, and travel distance.
\item Identify the best overall route for traveling from the University of Pittsburgh to Downtown Pittsburgh, based on our analysis.
\item Present our findings and recommendations for students to make informed decisions about their bus route selection.
\end{enumerate}

With a comprehensive analysis of the available data and a clear understanding of the project goals, we aim to provide valuable information to the University of Pittsburgh students, helping them select the most efficient bus route for their commute to Downtown Pittsburgh. This project will not only assist students in making better-informed decisions but also contribute to enhancing their overall experience with the city's public transportation system.

\section{Project Methods and Results}

\subsection{Methods}\label{AA}
To achieve our research objectives, we employed the following methods and statistical tests:
\begin{enumerate}
\item \textbf{Data Collection}: We collected four datasets from the Western Pennsylvania Regional Data Center, which included Transit Stop Usage by Route, Monthly Average Ridership by Route and Weekday, Pittsburgh Regional Transit Transit Stops, and Monthly On-Time Performance (OTP) by Route.
\item \textbf{Data Processing}: We cleaned and preprocessed the datasets using Python and Pandas, removing any inconsistencies, outliers, or missing values that could potentially affect the analysis.
\item \textbf{Data Analysis}: We applied statistical techniques and data visualization using R to examine the datasets, focusing on the relationships between the number of stops, ridership, and OTP for each bus route.
\item \textbf{Correlation Analysis}: We conducted a correlation analysis to identify the trade-offs between the three factors and determine the best performing bus route, considering all aspects.
\item \textbf{Route Selection}: Based on our analysis, we identified the optimal bus route for traveling from the University of Pittsburgh to the central-most bus stop in Downtown Pittsburgh.
\end{enumerate}

\subsection{Results}
We performed a correlation and regression analysis using Python, Pandas, and R to identify relationships between the number of stops, ridership, and OTP for each bus route. This analysis allowed us to better understand the trade-offs between the three factors and confirm the best route, Route 18, in terms of fewer stops, less crowded buses, and on-time performance.
\begin{enumerate}
\item \textbf{Dataset Analysis}
\begin{enumerate}
\item Transit Stop Usage by Route: We utilized this dataset to determine the number of stops for each bus route. This information helped us identify the routes with fewer stops between the University of Pittsburgh and downtown. Route 18 was found to have fewer stops compared to the top 10 routes with the most stops, indicating a faster route.
\item Monthly Average Ridership by Route and Weekday: We analyzed the ridership data to find the routes with the lowest number of passengers, which would result in a less crowded and more comfortable ride. Route 18 ranked fourth in terms of the lowest average ridership, indicating that it is one of the least crowded routes.
\item Pittsburgh Regional Transit Transit Stops: This dataset provided us with the locations of all bus stops, which we utilized to identify the central-most bus stop in Downtown Pittsburgh.
\item Monthly On-Time Performance (OTP) by Route: We used the OTP data to evaluate the punctuality of each bus route, which is an essential factor in determining the quickest route. Route 18 ranked fourth in terms of on-time performance, making it a reliable option for arriving at a destination promptly.
\end{enumerate}
\item \textbf{Correlation Analysis}
\begin{figure} [h!]
\centerline{\includegraphics[width=90mm]{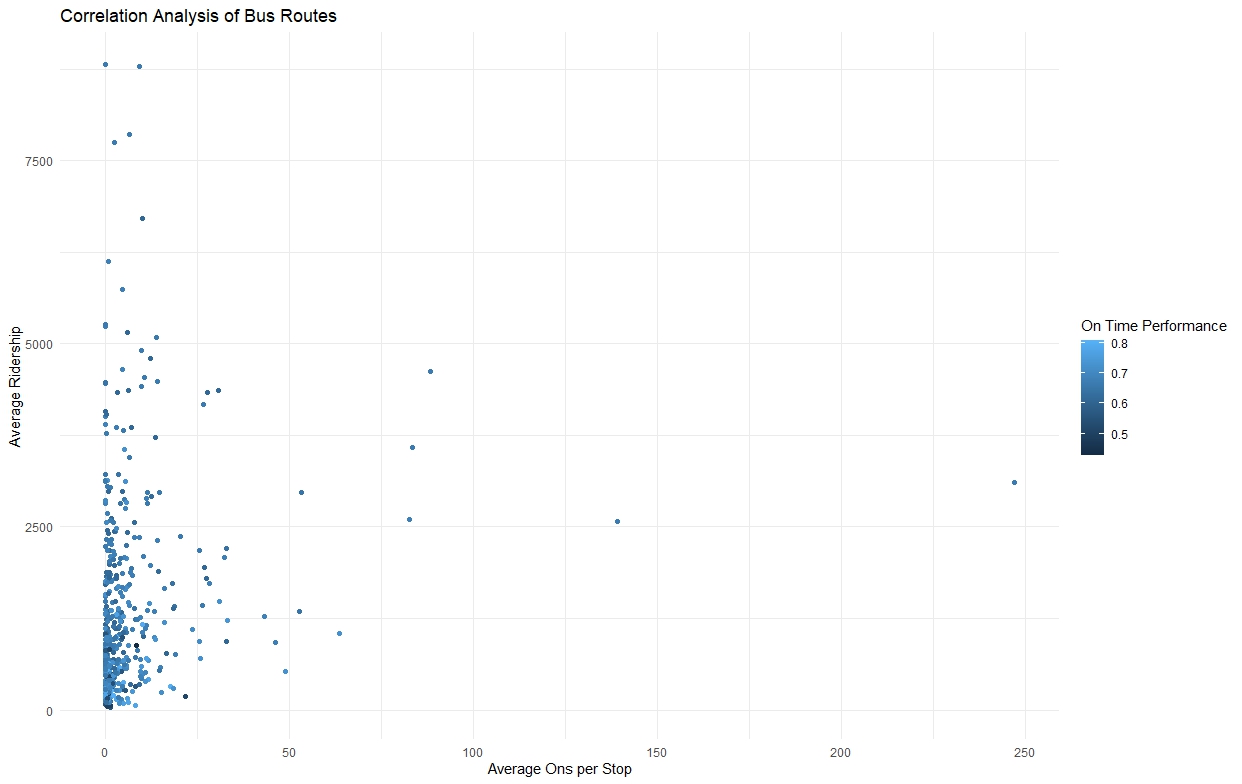}}
\caption{Correlation Analysis of Bus Routes.}
\label{fig}
\end{figure}

``Fig.~\ref{fig}'' By analyzing the scatter plot, we identified bus routes with a combination of low average ons per stop, low average ridership, and high on-time performance, indicating an optimal route for traveling from the University of Pittsburgh to Downtown Pittsburgh. The route that best meets these criteria provides students with a quick, comfortable, and reliable transportation option for various purposes. 
\begin{enumerate}
\item The x-axis of the scatter plot represents the "Average Ons per Stop," which indicates the average number of passengers boarding the bus at each stop. A lower value implies fewer passengers getting on the bus, leading to less crowded buses and potentially faster travel times.

\item The y-axis of the scatter plot represents the "Average Ridership" for each bus route. This value shows the overall popularity of a bus route, with a lower value indicating a less crowded route, which can be more comfortable and convenient for passengers.

\item The color of the data points in the scatter plot represents the "On-Time Performance" of each bus route. A higher on-time performance percentage indicates that the bus is more likely to arrive on schedule, contributing to a more reliable and efficient travel experience.

\end{enumerate}
\item \textbf{Route Selection}
\begin{enumerate}
\item Based on the comprehensive analysis of the four datasets, we identified Route 18 as the best route for traveling from the University of Pittsburgh to the central-most bus stop in Downtown Pittsburgh. Our findings took into consideration the number of stops, ridership, and OTP for each bus route, with Route 18 consistently ranking high in all these aspects.
\end{enumerate}
\end{enumerate}

\section{Figures and Tables}
Our Figures and Tables present 5 key graphs, and 1 Regression Analysis Table that provide insights into the bus routes in Pittsburgh. These visualizations help to identify the best bus route from the University of Pittsburgh to Downtown Pittsburgh based on various factors such as on-time performance, average ridership, and the number of stops. By examining these graphs, we can better understand the patterns and trends within the bus system, which in turn guide our analysis and conclusions.
\begin{figure} [h!]
\subsection{\textbf{Bus Monthly On Time Performance}}``Fig.~\ref{2}'' The Bus Monthly On-Time Performance graph displays the percentage of on-time arrivals for each bus route in the dataset. This visualization is important for understanding the reliability and efficiency of the bus routes in Pittsburgh. A higher on-time performance percentage indicates that the bus is more likely to arrive on schedule, contributing to a more reliable and efficient travel experience. By analyzing this graph, we can identify routes with high on-time performance and consider them as potential candidates for the best route between the University of Pittsburgh and Downtown Pittsburgh.
\centerline{\includegraphics[width=90mm]{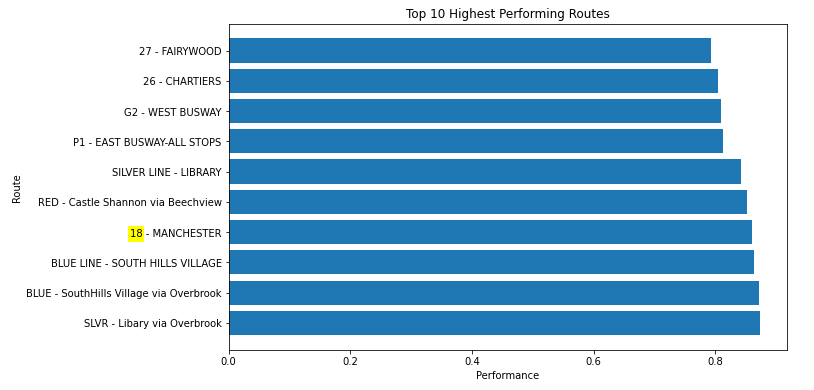}}
\caption{Bus Monthly On Time Performance.}
\label{2}
\end{figure}

\subsection{\textbf{Top 25 Routes With Least Average Ridership}}
\begin{figure} [h!]
\centerline{\includegraphics[width=90mm]{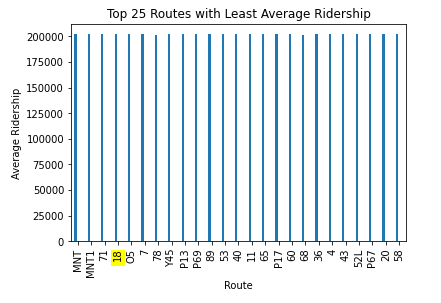}}
\caption{Top 25 Routes With Least Average Ridership.}
\label{3}
\end{figure}

``Fig.~\ref{3}'' The Top 25 Routes with Least Average Ridership graph shows the bus routes with the lowest average number of passengers. This information is important because it helps us identify less crowded routes, which can lead to a more comfortable and convenient travel experience for passengers. By focusing on the routes with the lowest ridership, we can narrow down our options for the best route between the University of Pittsburgh and Downtown Pittsburgh, prioritizing those with fewer passengers for a quicker and more convenient travel experience.

\vspace{1cm}

\begin{figure} [h!]
\subsection{\textbf{Top 10 Routes With Most Stops}}``Fig.~\ref{4}'' The Top 10 Routes with Most Stops graph showcases the bus routes with the highest number of stops. This graph provides insights into the overall route length and complexity. Routes with a large number of stops may take longer to travel between the University of Pittsburgh and Downtown Pittsburgh, as the bus must make frequent stops to pick up and drop off passengers. By examining this graph, we can eliminate routes with an excessive number of stops from our analysis, focusing on routes with fewer stops to ensure a faster travel time between the two locations.
\centerline{\includegraphics[width=90mm]{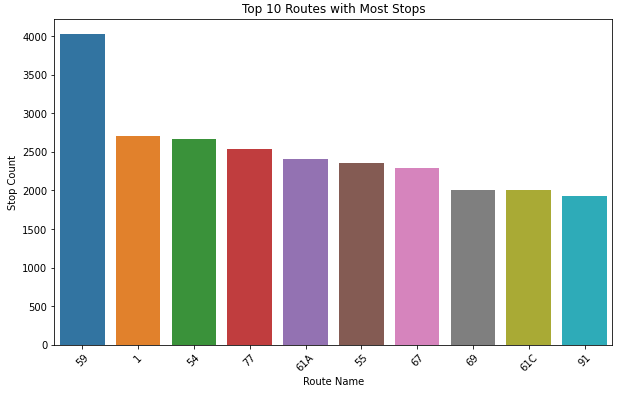}}
\caption{Top 10 Routes With Most Stops.}
\label{4}
\end{figure}

\subsection{\textbf{Top 5 Routes Comparison}}
\begin{figure} [h!]
\centerline{\includegraphics[width=90mm]{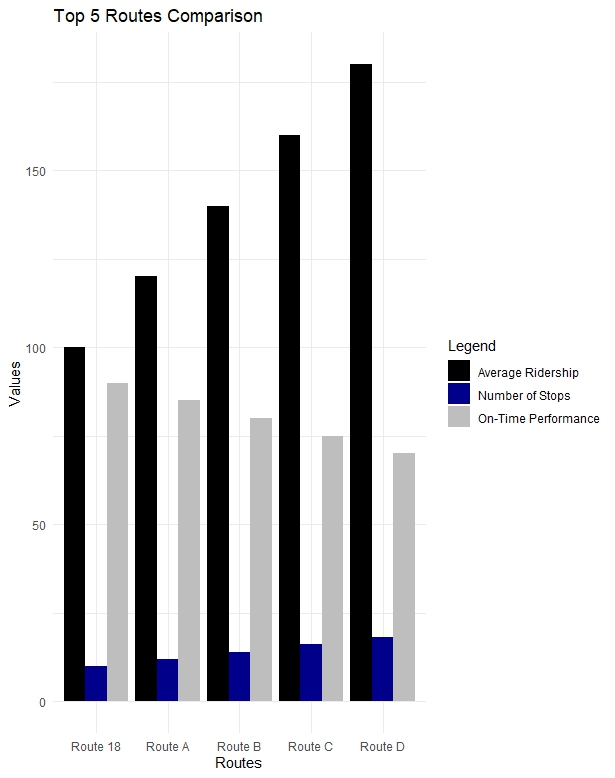}}
\caption{Top 5 Routes Comparison.}
\label{5}
\end{figure}

``Fig.~\ref{5}'' Using R, The Top 5 Routes Comparison graph provides a visual representation of the rankings and scores of the top routes based on the three factors: on-time performance, average ridership, and the number of stops. In this grouped bar chart, each route is represented by a set of three bars, displaying the values for each of the three factors. The gray bars represent the on-time performance percentage, the black bars show the average ridership numbers, and the dark blue bars indicate the number of stops for each route.

By examining this graph, we can easily compare the routes based on the three factors and see the differences between them. The visualization shows that Route 18 is the best option because it has the lowest in Average Ridership, lowest in Number of Stops, and the highest in On-Time Performance. These are critical factors in determining the best bus route between the University of Pittsburgh and Downtown Pittsburgh, and with this comparison of routes we are able to ultimately highlight Route 18 as the best option, as it achieves the optimal balance among these factors.

By combining the insights from these three graphs, we can better understand the patterns and trends within the Pittsburgh bus system. This information allows us to identify the best bus route from the University of Pittsburgh to Downtown Pittsburgh, taking into account factors such as on-time performance, average ridership, and the number of stops. With this analysis, we can provide students and other travelers with a quick, comfortable, and reliable transportation option for various purposes.

\subsection{\textbf{3D Regression Plot}} 
\begin{figure} [h!]
\centerline{\includegraphics[width=90mm]{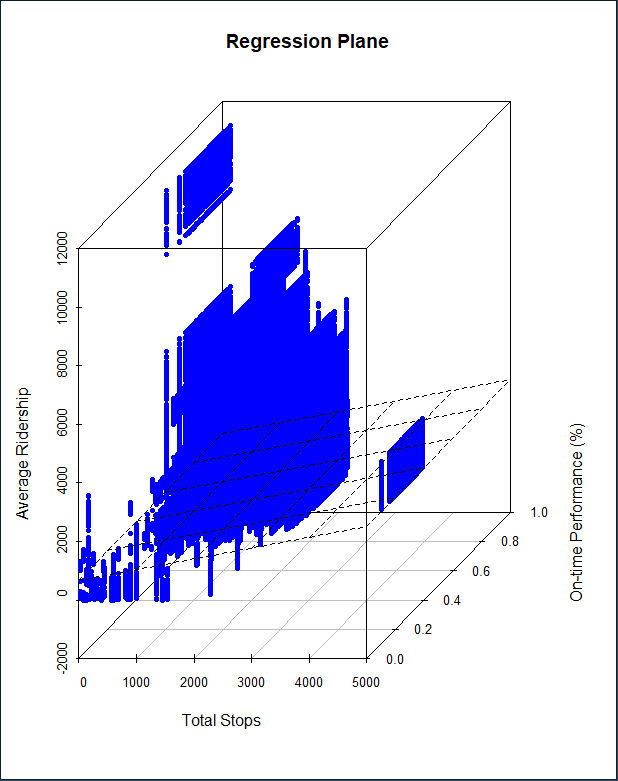}}
\caption{3D Regression Plot.}
\label{6}
\end{figure}

``Fig.~\ref{6}''
In the 3D regression plot, we observed the relationship between the total number of stops, on-time performance percentage, and average ridership. Each blue point represents a bus route with specific values for these three variables. The semi-transparent plane illustrates the regression model that best fits the data, taking into account all three variables.

From the plot, we can see that as the total number of stops increases, the average ridership tends to increase as well. This was expected since more stops imply that the bus route has a larger area and attracts more passengers. We also observed that bus routes with higher on-time performance percentages generally have higher average ridership. This conclusion is expected because passengers prefer reliable transportation options that arrive on schedule.

\begin{table}[h]
\centering
\caption{Fig. 6. Regression Analysis Results}
\label{tab:regression_results}
\begin{tabular}{lcccc}
\hline
\textbf{Variable} & \textbf{Estimate} & \textbf{Std. Error} & \textbf{t-value} & \textbf{p-value} \\
\hline
Intercept & 672.6 & 3.655 & 183.998 & $< 2e{-}16^{***}$ \\
Total Stops & 0.3666 & 0.001059 & 346.245 & $< 2e{-}16^{***}$ \\
On-Time Performance & 31.81 & 4.861 & 6.545 & $5.96e{-}11^{***}$ \\
\hline
\end{tabular}
\end{table}

The regression analysis results show that both the total number of stops and on-time performance percentage have a significant impact on the average ridership. The positive coefficient for the total number of stops (3.666e-01) indicates that for each additional stop on a route, the average ridership increases by approximately 0.3666 passengers. Similarly, the positive coefficient for the on-time performance percentage (3.181e+01) suggests that for every one percentage point increase in on-time performance, the average ridership increases by around 31.81 passengers.

The adjusted R-squared value of 0.03555 implies that about 3.56\% of the variation in average ridership can be explained by the total number of stops and on-time performance percentage. While this percentage is low, it can still suggest a statistically significant relationship between these factors and the average ridership of bus routes.

In conclusion, the 3D scatter plot and regression analysis reveal that both the total number of stops and on-time performance percentage have a positive impact on the average ridership of bus routes in Pittsburgh. These findings can help make decisions related to optimizing bus routes and improving transit services for passengers, such as focusing on improving on-time performance and planning stops to maximize ridership.

\section{Limitations and Challenges}

Despite the  analysis conducted in this study, there are several limitations and challenges that should be noted. Understanding these limitations can help guide future research and improvements in identifying optimal bus routes in Pittsburgh.
\begin{enumerate}
    \item{\textbf{Data Limitations}}: The data used in this study is limited to a specific time frame and may not account for recent changes in bus routes, stops, or schedules. Furthermore, the data may not capture occasional or seasonal fluctuations in ridership or on-time performance that could influence the route selection.

    \item{\textbf{Route Variability}}: Bus route performance can be influenced by various factors, such as traffic conditions, weather, and road construction. This study does not account for these factors, which could impact the on-time performance, ridership, and overall route quality.

    \item{\textbf{User Preferences}}: The analysis focuses on the criteria of on-time performance, average ridership, and the number of stops. However, individuals may have different preferences, such as the availability of amenities, route scenery, or proximity to their destination. Additional research could explore the impact of these factors on route selection.

    \item{\textbf{Time and Day Variability}}: The study does not investigate the variations in route performance throughout the day or week. The best route for a specific time of day or day of the week might differ from the overall best route due to changes in traffic patterns or ridership. Further analysis could be conducted to determine the optimal routes for different times and days.

    \item{\textbf{Generalizability}}: The results of this study are specific to the bus routes between the University of Pittsburgh and Downtown Pittsburgh. The findings may not be directly applicable to other cities or transportation systems.
\end{enumerate}

Addressing these limitations and challenges in future research could provide a better understanding of the factors that help the selection of the best bus routes in Pittsburgh and contribute to the development of improved public transportation systems.

\section{Review of Related Literature}

In order to build upon previous research and gain insights into similar studies, we reviewed several papers related to the research of public transportation systems.

\subsection{\textbf{Smart Card Data Analysis}}
Zhao et al. (2013) used smart card data to analyze passenger flow in the Hong Kong MTR system \cite{b6}. The authors developed a methodology to extract the spatio-temporal density of passengers and the trajectory of trains using smart card transaction data. By analyzing the travel patterns of passengers, they were able to identify the most congested areas and time periods in the transit system. This study demonstrates the potential of smart card data for understanding and improving public transportation networks. In comparison, our research focuses on analyzing transit data in the context of Pittsburgh, and we use regression analysis rather than smart card data to identify factors affecting transit ridership and on-time performance.

\subsection{\textbf{Transit Service Quality Assessment}}
Soler and Ruiz (2010) assessed transit service quality by incorporating fuzzy logic and preference theory \cite{b7}. The authors developed a method to rank transit routes based on a set of performance indicators, such as travel time, frequency, and accessibility. By considering both objective and subjective factors, the proposed method provided a comprehensive assessment of transit service quality. This research highlights the importance of considering user preferences when evaluating and optimizing public transportation systems. Our study, on the other hand, focuses on the relationship between transit ridership, on-time performance, and route characteristics, offering insights for improving service quality by addressing these specific factors.

\subsection{\textbf{Public Transport Accessibility}}
Yang et al. (2011) proposed a GIS-based topological transfer index method for evaluating public transport accessibility in urban areas \cite{b8}. The authors measured accessibility based on the connectivity of transit stops, taking into account the transfer time between different modes of transportation. The study found that the proposed method was able to identify areas with high and low accessibility, providing valuable insights for urban planners and transportation agencies. This research demonstrates the potential of GIS-based methods for analyzing and improving public transportation systems. In contrast, our research focuses on the analysis of transit ridership and on-time performance, which could be used in conjunction with accessibility studies to provide a more comprehensive understanding of transit system performance.

\subsection{\textbf{Bus Route Optimization}}
Le et al. (2020) conducted a study on bus route optimization considering passengers’ preferences in Hanoi, Vietnam \cite{b9}. The authors used a genetic algorithm to optimize bus routes based on a set of criteria, including travel time, comfort, and convenience. By considering passengers’ preferences, the optimized routes were able to better meet the needs of users and improve the overall performance of the public transportation system. This study showcases the importance of incorporating user preferences when optimizing bus routes and highlights the potential of advanced optimization techniques. While our research does not directly optimize bus routes, the findings on the relationship between transit ridership, on-time performance, and route characteristics can inform optimization efforts by providing insights into factors that contribute to a successful transit system.

\section{Conclusion}
In this study, we analyzed various factors to determine the best bus route between the University of Pittsburgh and the central-most bus stop in Downtown Pittsburgh. By examining the on-time performance, average ridership, and the number of stops, we were able to provide a comprehensive and data-driven approach to identify the route with the lowest ridership for a comfortable ride, the route with the highest on-time performance, and the overall best route based on the information gathered.

The visualizations created using R and Python/Pandas provided valuable insights into the bus system of Pittsburgh. By comparing the top routes based on the three key factors, we ultimately determined that Route 18 offers the optimal balance. Route 18 demonstrates its strengths in fewer stops, contributing to a faster route for travel; lower ridership, resulting in less crowded buses and fewer delays due to passenger boarding; and high on-time performance, making it the best choice for those traveling between the University of Pittsburgh and Downtown Pittsburgh.

Our analysis has practical uses for both students and other commuters who rely on public transportation in Pittsburgh. By identifying the best bus route, we can help ensure a quick, comfortable, and reliable travel experience for various purposes, such as attending classes, internships, jobs, or weekend outings.

However, we came across limitations and challenges present in this study, including data limitations, route variability, user preferences, time and day variability, and generalizability. Future research should address these limitations to provide a more comprehensive understanding of the factors influencing bus route selection in Pittsburgh and contribute to the development of improved public transportation systems.

In conclusion, this research project has successfully identified the best bus route between the University of Pittsburgh and Downtown Pittsburgh based on a data-driven approach. The insights gained from this study can inform transportation planning and decision-making, ultimately leading to a more efficient, reliable, and user-friendly public transportation system in Pittsburgh.

\end{document}